\documentclass[prb,aps, twocolumn, superscriptaddress,longbibliography]{revtex4-1}
\usepackage{graphicx}
\usepackage{wasysym}
\usepackage{amsfonts}
\usepackage{bm}
\usepackage{enumerate}
\usepackage{color}
\usepackage[resetlabels]{multibib}

\newcites{sec}{Supplementary References}
\usepackage{times}
\newcommand{\beq}{\begin{equation}}
\newcommand{\eeq}{\end{equation}}
\newcommand{\bea}{\begin{eqnarray}}
\newcommand{\eea}{\end{eqnarray}}

\newcommand{\bg}{b^{\dg}}

\mathchardef\nss="711B

\def\zhat{{\hat z}}

\def\xhat{\hat{\boldsymbol{x}}}
\def\yhat{\hat{\boldsymbol{y}}}
\def\zhat{\hat{\boldsymbol{z}}}

\def\ket#1{{\left|#1\right\rangle}}
\def\bra#1{{\left\langle #1 \right|}}

\def\br{\boldsymbol{r}}
\def\bk{\boldsymbol{k}}

\def\btau{{\bm \tau}}

\def\bg{{g}}
\def\bh{{h}}
\def\vk{\boldsymbol{k}}

\def\nss{\mathcal{S}}
\def\va{\boldsymbol{a}}

\def\vr{\boldsymbol{r}}

\def\plz{\boldsymbol{\mathcal{P}}}

\def\be{\begin{eqnarray}}
\def\ee{\end{eqnarray}}

\newlength{\myL}

\begin{document}
\title{Topological Order and Absence of Band Insulators at Integer Filling in Non-Symmorphic Crystals}
\author{Siddharth A. Parameswaran}
\email{sidp@berkeley.edu}
\affiliation{Department of Physics, University of California, Berkeley, CA 94720, USA}
\author{Ari M. Turner}
\affiliation{Institute for Theoretical Physics, University of Amsterdam, Science Park 904, P.O. Box 94485, 1090 GL Amsterdam, The Netherlands}
\author{Daniel P. Arovas}
\affiliation{Department of Physics, University of California San Diego, La Jolla,  92093, USA}
\author{Ashvin Vishwanath}
\affiliation{Department of Physics, University of California, Berkeley, CA 94720, USA}
\affiliation{Materials Science Division, Lawrence Berkeley National Laboratories, Berkeley, CA 94720}

\date{\today}
\begin{abstract}
 Band insulators appear in a crystalline system only
when the filling -- the number of electrons per unit cell and spin
projection -- is an integer. At fractional filling, an insulating phase that
preserves all symmetries is a Mott insulator, i.e. it is either gapless or, if gapped, displays
fractionalized excitations and topological order. We raise the inverse
question -- at an integer filling is a band insulator always possible? Here
we show that lattice symmetries may forbid a band insulator even at
certain integer fillings, if the crystal is non-symmorphic -- a property shared by a majority of
 three-dimensional crystal structures. In these cases, one may infer the existence of 
 topological order if the ground state is gapped and fully symmetric. This is demonstrated using a
non-perturbative flux threading argument, which has immediate applications
to quantum spin systems and bosonic insulators in addition to
electronic band structures in the absence of spin-orbit interactions.
\end{abstract}
\maketitle
Fundamental to the study of crystalline materials is the existence of band insulators, and the fact that they only occur when the electronic filling is commensurate with the lattice. Defining the filling as one half the number of electrons in a unit cell 
-- which accounts for spin degeneracy in the absence of spin-orbit coupling -- band insulators occur {\em only}  if the filling is an integer.   At fractional filling a material must be metallic within the free electron approach. However, interactions could lead to a distinct insulating phase.
For fermionic systems, this would have to be a Mott insulator, strictly defined as an insulator that breaks no symmetries but is distinct, {\it i.e.} separated by a phase transition, from a band insulator. In fact, a far stronger result is true: even interactions cannot
cause a system at
fractional
filling, whether of bosons or fermions, to enter  a trivial insulating state.
There are only two choices if
all symmetries are to be preserved: the system must remain gapless, or develop
topological order, which we take here to mean 
a ground state degeneracy that is not associated to a broken symmetry,  as exemplified by fractional quantum Hall states and
gapped quantum spin liquids. {These gapped states feature long range quantum entanglement, leading to excitations with anyonic statistics and ground state degeneracy that is sensitive to the spatial topology.  }
A concrete example is  the the one-dimensional ($d=1$) Hubbard model at half filling, with repulsive interactions whose ground state is insulating. The spin excitations in the insulator form a gapless Luttinger liquid, as expected of a fully symmetric state in $d=1$, where a gapped state with topological order is forbidden.
This result was demonstrated for arbitrary spatial dimension by Hastings\cite{Hastings:2004p1,Hastings:2005p1} and Oshikawa,\cite{Oshikawa:2000p1} by extending the $d=1$ Lieb-Schultz-Mattis\cite{Lieb:1961p1} theorem to $d>1$ using a beautiful `flux-threading' argument.\cite{AltmanAuerbach, OshikawaYamanakaAffleck}

However, surprisingly little is known about the inverse question: given an integer filling, is it always possible to find a `trivial'  insulating state, like a band insulator, that is not required to have these exotic properties?
Here we show
 that for many crystals, trivial
  insulators can be forbidden {even though the filling is an integer}.
For bosons or fermions, if the insulating states preserve symmetry, they must either be topologically ordered or host gapless excitations.  We demonstrate a simple crystallographic criterion for this to hold: namely, that the space group of the crystal is  {\it non-symmorphic}.
 A  crystal is non-symmorphic if there is no choice of origin about which
 all its symmetries can be decomposed into a product of
  a lattice translation and a point group symmetry element.
 Our key result is succinctly stated: {\em There are integer fillings at which insulators with unique ground states are impossible if the space group is non-symmorphic.}
  For such crystals we show that there is a minimal integer filling, strictly greater than unity, at which trivial insulators are allowed. We call this the  non-symmorphic rank, denoted by  $\nss>1$. Trivial insulators exist only at fillings which are integer multiples of $\nss$.

 We may rationalize this unexpected result by observing  that in non-symmorphic crystals, a fractional lattice translation acts in concert with another transformation to leave the crystal invariant. This, loosely speaking, renders an integer filling equivalent to a fractional one, thus forbidding a trivial insulator.  Similar conclusions can be drawn for any system where it is possible to identify a conserved charge. These include spin systems with a conserved spin component, and lattice bosons whose number is conserved. Note, 157 of the 230 distinct space groups in three dimensions are non-symmorphic and have $\nss>1$. These include the ubiquitous  hexagonal close-packed (hcp) structure and the diamond lattice space group relevant to many semiconductors and pyrochlore materials. Table~\ref{tab:ranks} has further examples. The implications therefore are manifold, a few applications will be discussed below.

Before we move to the main results, it is worth making two observations. First, we make no assumptions beyond lattice symmetry. In particular we do not limit ourselves to tight-binding models where additional restrictions may emerge.\cite{YaoKivelson,HoneycombVoronoi}
 Indeed, the construction of trivial interacting insulating states in those cases relies precisely on the fact that they do not violate the conditions we establish here. Second, when we discuss electronic band structures, we assume that the electron's spin is purely a spectator, {\it i.e.} there is no spin-orbit coupling. This allows us to address each spin species independently, as though the electrons were `spinless'. Moreover, we will connect our key results to extinguished Bragg peaks,  which assumes a `scalar' coupling between electrons and the crystal lattice. Incorporating electronic spin-orbit coupling is an important extension which we leave to future work.

\section{Applications}
\noindent{\it (i) Band Theory.--} The fact that band insulators are forbidden unless the filling is a multiple of the non-symmorphic rank $\nss$, strongly constrains the  structure  of the bulk energy dispersion: it is impossible to `detach' a set of fewer than $\nss$ bands so that they touch no other bands, without breaking the crystal symmetry. Non-symmorphic space groups often also describe the symmetry of photonic crystals,\cite{YablonovichPhotCryst,SajeevJohnPhotCryst}  hence their photonic band structures obey similar constraints.
 For instance, the hcp structure has $\nss=2$; we show a tight-binding band structure for the hcp lattice in Fig.~\ref{fig:hcp} where the enforced contacts are explicit.
Although the subject of band touchings in crystals has a long history,\cite{HerringMonodromy, HerringTRS} and the ubiquity of such degeneracies in non-symmorphic crystals has been noted,\cite{Konig:1997p1,MichelZakNSConnectivity} the connection to a minimum filling for band insulating behavior has not been made previously. More importantly, these prior results  apply only to noninteracting systems, in contrast to the non-perturbative approach taken here, which allows us to determine the nature of the interacting insulating ground state.

\noindent{\it (ii) Spin Systems and Bosonic Insulators.--} A parallel set of conclusions can be drawn for spin systems, where the filling is related to the total magnetization, and an  insulator now corresponds to a phase with a spin gap.  We demand at least a U(1) spin rotation symmetry, although the conclusions also apply to a larger symmetry such as SU(2), so long as it contains a U(1) subgroup. The analog of the band insulator is a trivial paramagnet, that has gapped excitations, and has neither conventional nor topological order. We conclude that such a trivial paramagnet is disallowed in an SU(2) symmetric spin-1/2 system on the diamond lattice, which is a common non-symmorphic lattice with  $\nss=2$. In contrast, the pyrochlore lattice has the same space group but twice as many sites per unit cell as diamond, and a trivial quantum paramagnet is not forbidden by our arguments. Another application is to magnetization plateaus in an applied Zeeman field.
For example, half magnetization plateaus of  spin-1/2 moments on the two-dimensional non-symmorphic Shastry-Sutherland lattice\cite{ShastrySutherland} (SSL) 
cannot be trivial paramagnets. Applications to  SrCu$_2$(BO$_3$)$_2$ (SCBO), a material realizing the SSL where a half magnetization plateau is observed,\cite{SebastianSSLHalfMagPlateau,SSLHalfMagExpt} will be discussed below. Yet another application is to bosonic Mott insulators. Our arguments demonstrate that on non-symmorphic lattices, Mott insulators at fillings that are not a multiple of $\mathcal{S}$ must be topologically ordered, if they are gapped and respect all symmetries.

\section{Flux Threading Argument and Non-Symmorphic Rank} In the balance of this paper we outline our argument, and demonstrate its use in specific examples by  substantiating the claims made above. For clarity and to fix notation we digress briefly to review some relevant crystallography. We will consider crystalline systems with a given space group, $\mathcal{G}$. This has two ingredients: (i) the subgroup of translations $\mathcal{T}$, generated by the set of {\it primitive} translations $\hat{t}_{\va_i}:\br \rightarrow \br +\va_i$, $i=1,2,3$, where the $\va_i$ generate a specific Bravais lattice, and (ii) a point group, $\mathcal{P}$ consisting of discrete rotations, inversions, and reflections. By combining the 14 Bravais lattices with the 32 crystallographic point groups, we obtain the 230 space groups. Of these, $73$ are  {symmorphic}: there exists at least one point that is left invariant by all the symmetries, up to translations by a lattice vector. The remaining 157 space groups for which no such point exists, are non-symmorphic. Intuitively, a non-symmorphic space group contains one or more {\it essential} non-symmorphic operations
 (`glide mirrors' or `screw rotations') that combine a point-group operation with a fractional lattice translation,  that {\it cannot} by any change of origin be rewritten as the product of a point-group operation and an ordinary translation. The latter caveat is important to distinguish these from `trivial' glides/screws which can occur even in symmorphic crystals.\cite{Konig:1999p1}
 There are however two `exceptional' space groups, which are non-symmorphic despite the absence of essential glide or screw operations.\cite{SuppMat}

In the remainder, we focus on the 155 non-symmorphic groups which have glide mirrors or screw rotations. Under a non-symmorphic symmetry $\hat{G}$ consisting of a point group operation $g$ and non-lattice translation $\btau$, the Fourier components of a scalar field (such as the density) transform as $n_{\bk} \rightarrow n_{g\bk} e^{i\btau\cdot\bk}$.
 If $g\bk =\bk$ and $\btau\cdot\bk$ is not an integer multiple of $2\pi$, $n_{\bk} =0$: the associated Fourier component vanishes. An essential glide or screw always has an infinite set of such reciprocal lattice vectors $\bk$.  Thus non-symmorphicity (barring two exceptions) has a dramatic experimental manifestation: there are systematic absences in the diffraction pattern where the (scalar) Bragg intensity vanishes.

We  assume that the systems we study are described by a Hamiltonian $\hat{H}$  that preserves all the symmetries of $\mathcal{G}$ and in addition that there is a conserved U(1) charge, $\hat{\mathcal{Q}}$ (assumed to take a fixed integer eigenvalue and thus replaceable by a $c$-number throughout) 
 with $[\hat{H},\hat{\mathcal{Q}}] =0$. 
 We make no assumptions as to the origins of the conserved charge, so for instance the systems we consider could be built out of (i) spinless fermions or bosons, where $\mathcal{Q}$ is just the conserved particle number; (ii) spinful fermions with SU(2) spin symmetry, in which case $\mathcal{Q}$ is one-half the total fermion number (since the two spin components may be treated independently); or (iii) lattice spins with (at least) U(1) spin rotational invariance, in which case we may take 
the charge on lattice site $\br$ to be $S+ \hat{S}^z_{\br}$  where $S, \hat{S}^z_{\br}$ are its total spin and magnetization, and define $\mathcal{Q}$ accordingly. Considering a finite system with $\mathcal{N}_c = N^3$ unit cells, we may then define the {\it filling} to be the charge per unit cell, {\it i.e.} $\nu =  \mathcal{Q}/\mathcal{N}_c$, which will be held fixed in the thermodynamic limit.
 We impose periodic boundary conditions that identify  ${\boldsymbol{r}}$ and $ {\boldsymbol{r}}+N\va_i$.

Our argument is based upon the response of the system to the  insertion of a gauge flux that couples minimally to the conserved charge $\mathcal{Q}$. We first give a heuristic argument, and then a more formal one. The strategy is as follows: we wish to show that if the system is an insulator, the ground state cannot be unique in the thermodynamic limit. Hence the system cannot be
 a trivial insulator that respects all symmetries, which should have a unique ground state. A degenerate ground state implies either  (i) a broken symmetry or (ii) a gapless phase or (iii) topological order. That is, it is topologically ordered  {if it is gapped and breaks no symmetries}. To show this, we begin with a ground state $|\Psi\rangle$ and thread a flux quantum through a periodic direction, which, by gauge invariance, returns us to the original Hamiltonian. This procedure produces an eigenstate $|\tilde{\Psi}\rangle$. Earlier work\cite{PhysRevLett.90.236401,PhysRevB.70.245118,Hastings:2004p1,Hastings:2005p1} has argued that for an insulator, $|\tilde{\Psi}\rangle$ must be a `low energy' state, {\it i.e.} its energy approaches that of the ground state in the thermodynamic limit.
 {Although rigorous energy bounds can only be given for a different -- but gauge-equivalent -- flux insertion, for pedagogical reasons we keep a simpler choice with the understanding that the arguments of Ref.~\onlinecite{Hastings:2005p1} 
 can be applied, {\it mutatis mutandis}, to non-symmorphic symmetries.}
The key step is to show that $|\tilde{\Psi}\rangle$ is distinct from  $|\Psi\rangle$, which would then establish ground state degeneracy. In the case of fractional filling, these states differ in crystal momentum.\cite{PhysRevLett.90.236401,PhysRevB.70.245118,Hastings:2004p1,Hastings:2005p1} For integer filling, crystal momentum fails to differentiate between them. However, we show that on non-symmorphic lattices, one can still distinguish these states using the quantum numbers of the non-symmorphic operations.
\begin{figure}
\includegraphics[width=\columnwidth]{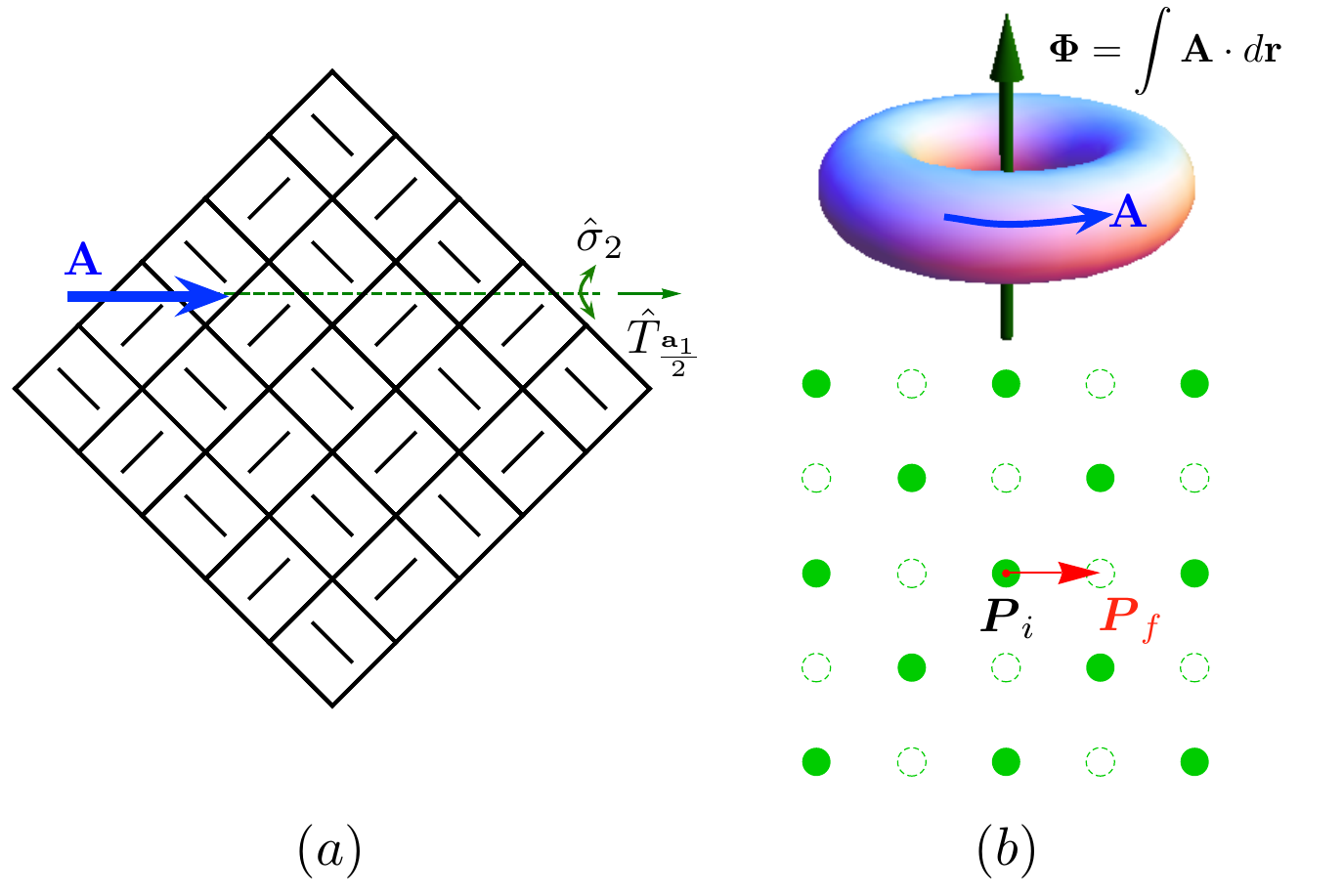}\caption{\label{fig:p4g}Flux Threading on a Non-symmorphic Lattice. ({\bf a}) A 2D non-symmorphic lattice, with p4g space group. An essential glide symmetry is shown,  consisting of a mirror ($\hat{\sigma}_2$) and half-translation ($\hat{T}_{\va_1/2}$). ({\bf b}) Flux threading changes ground state momentum from $\boldsymbol{P}_i$ to $\boldsymbol{P}_f$. At odd integer filling, $\boldsymbol{P}_f$ is an extinguished reciprocal lattice vector, denoted by empty circles, implying a distinct ground state. At even filling, $\boldsymbol{P}_f$ lands on an allowed reciprocal lattice vector, allowing for a unique ground state.} \end{figure}

Consider threading flux through the system by introducing a vector potential $\boldsymbol{A} =  \vk/N$ , where $\vk$ is a reciprocal lattice vector; this is the most general vector potential that is `pure gauge' so that $e^{i \int_{\mathcal{C}} \boldsymbol{A}\cdot d\br} = 1$ for any loop $\mathcal{C}$ that threads the system. The change in energy upon inserting flux is proportional to the total current and is thus bounded in an insulator, where this procedure produces a state degenerate with the ground state in the thermodynamic limit. If the flux threading changes the quantum numbers, the final state is distinct from the initial one and thus the ground state is degenerate. { We use units where $\hbar=e=1$, in which the flux quantum $\phi_0=2\pi$. In the case of spinful fermions, we consider coupling to a single spin species, justified by the fact that spin is assumed to be a spectator.}

In order to determine the change in quantum numbers of the symmetry operators, it is useful to first ignore the lattice potential, in which case it is straightforward to identify the change of the center-of-mass momentum by computing the force imparted to each charge as the flux is adiabatically switched on. Faraday's law gives $\boldsymbol{F}_i = \int \dot{\boldsymbol{A}}\cdot d\br$, so that
\be
\Delta\boldsymbol{P} =\sum_{i=1}^{\mathcal{Q}} \int \boldsymbol{F}_i dt =   \frac{\mathcal{Q}}{N} \vk =  N^2\nu \vk
\ee
We now reintroduce the lattice. {If $\Delta {\bf P}$ is not in the reciprocal lattice, then it is observable even within the reduced symmetry of the crystal, and we have succeeded in producing a distinct degenerate ground state.  For a fractional filling $\nu=p/q$, it is clear that this is the case so long as we choose $N$ relatively prime to $q$.  In other words, the state following flux insertion  has a distinct crystal momentum, which means that the quantum number associated with translational symmetry has changed.} This is the essence of the Hastings-Oshikawa argument. At integer filling, it is clear that no choice of $N$ allows us to distinguish the initial and final states on the basis of lattice translational symmetry.

However, in a non-symmorphic crystal, if we can choose $\Delta\boldsymbol{P}$ to correspond to an extinguished Bragg peak, the initial and final states will have different quantum numbers for the non-symmorphic operation responsible for that extinction. Such a choice is always possible at unit filling, implying that an insulator with a unique ground state is forbidden. Intuitively, the absence of Bragg peaks at special crystal momenta indicates that they can serve as a good quantum number to distinguish initial and final states, in the presence of a crystalline potential. In general, as long as $\nu$ is not a multiple of the {\it non-symmorphic rank} $\mathcal{S}$, a unique ground state is forbidden.\cite{SuppMat}  This result may also be understood by the exercise of trying to isolate a single Bloch band in a non-symmorphic crystal without breaking symmetry. Were such a band to exist, its exponentially localized Wannier orbitals would define centers of electronic charge in each unit cell. One now encounters an obstruction:  it is impossible to define a charge center that is invariant, modulo translations, under all crystal symmetries.

We now bolster this intuitive picture with a more formal argument. Consider a crystal which has a non-symmorphic space group $\mathcal{G}$, which contains a non-symmorphic operation $\hat{G}$. This comprises a point group operation $g$ followed by a fractional translation $\btau$ in a direction left invariant by $g$, {\it i.e.}
\be\hat{G}: \vr \rightarrow g\vr + \btau\ee
where $\btau$ is {\it not} a lattice vector, and $g\btau =\btau$.
We begin with a ground state $\ket{\Psi}$, and assume it is an eigenstate of all the crystal symmetries, including $\hat{G}$, {\it i.e.}
\be
\hat{G}\ket{\Psi} = e^{i \theta} \ket{\Psi}
\ee
We consider the smallest reciprocal lattice vector $\vk$    left invariant by $g$, so that $g\vk = \vk$ and $\vk$ generates the invariant sublattice along $\hat{\vk}$.   We now thread flux by introducing a vector potential $\boldsymbol{A} = \bk/N$ as before, in the process of which $\ket{\Psi}$ evolves to a state $\ket{\Psi'}$ that is degenerate with it. To compare $\ket{\Psi'}$ to $\ket{\Psi}$, we must return to the original gauge, which can be accomplished by the unitary transformation $\ket{\Psi'} \rightarrow \hat{U}_{\vk}\ket{\Psi'}\equiv|\tilde{\Psi}\rangle $,  where
\be
\hat{U}_{\vk} = \exp\left\{{ \frac{i}{N} \int d^d{r}\, \vk\cdot\vr \hat{\rho}(\br) }\right\}
\ee
removes the  inserted flux, and $\hat{\rho}(\br)$ is the density operator corresponding to the conserved charge $\hat{\mathcal Q} $.
Since  $\boldsymbol{A}$ is left invariant
 by  $\hat{G}$, threading flux does not alter $\hat{G}$ eigenvalues, so $\ket{\Psi}$ and $\ket{\Psi'}$ have the same quantum number under $\hat{G}$; however, on acting with $\hat{U}_{\vk}$, the eigenvalue changes, as can be computed from the equation:
\be\label{eq:FCR}
\hat{G}^{-1} \hat{U}_{\vk}\hat{G} =  \hat{U}_{\vk} e^{2\pi i \Phi_g(\vk){\mathcal{Q}}/N}\ee
where  we have defined the phase factor $ \Phi_g(\vk) = \btau\cdot \vk/{2\pi}$, and ${\mathcal Q}=\nu N^3$ is the total charge.
 It may be readily verified that since $g\vk=\vk$, $\Phi_g(\vk)$ is unchanged by a shift in real-space origin. For a non-symmorphic symmetry operation $\hat{G}$,  this phase  $\Phi_g(\vk)$  must be a fraction. This follows since $\btau$ is a fractional translation. (If a lattice translation had the same projection onto $\vk$ as $\btau$,  this would yield an integer phase factor.\cite{Konig:1999p1} However, this would render the screw/glide removable  {\it i.e.} reduced to point group element$\times$translation by change of origin.)  Thus, for  $\hat{G}$  non-symmorphic, $\Phi_g(\vk) = p/\nss_G$, with $p,\nss_G$ relatively prime.
From (\ref{eq:FCR}) we conclude that $\ket{\Psi}$ and $|\tilde{\Psi}\rangle$
 have distinct $\hat{G}$ eigenvalues whenever $\Phi_g(\vk) \mathcal{Q}/N  =p  N^2\nu/\nss_G $ is a fraction. Since we may always choose $N$ relatively prime to the $\nss_G$, the result of flux insertion is a state distinguished from the original state by its $\hat{G}$ eigenvalue, unless the filling is a multiple of $\nss_G$. For a glide $\nss_G=2$, while for a screw $\nss_G$ is the number of times it must be applied before it becomes removable.\cite{SuppMat}

{\it Non-Symmorphic Rank.--}  Recall that trivial insulators are only allowed at filling that are multiples of an integer $\nss$, which we call the non-symmorphic rank. Each non-symmorphic operation $\hat{G}$ is individually associated with an integer $\nss_G$ which leads to a degeneracy unless it divides the filling $\nu$.
Hence the non-symmorphic rank $\nss$  is divisible by the least common multiple of the $\nss_G$.
 A tighter bound on the non-symmorphic rank, although subtle to prove\cite{SuppMat} is easy to state: it is
the {smallest integer} $\nss$ such that some point is invariant under all the elements
of $\mathcal{G}$ up to ${1}/{\nss}$ times Bravais lattice vectors. 

\begin{figure}
\includegraphics[width=\columnwidth]{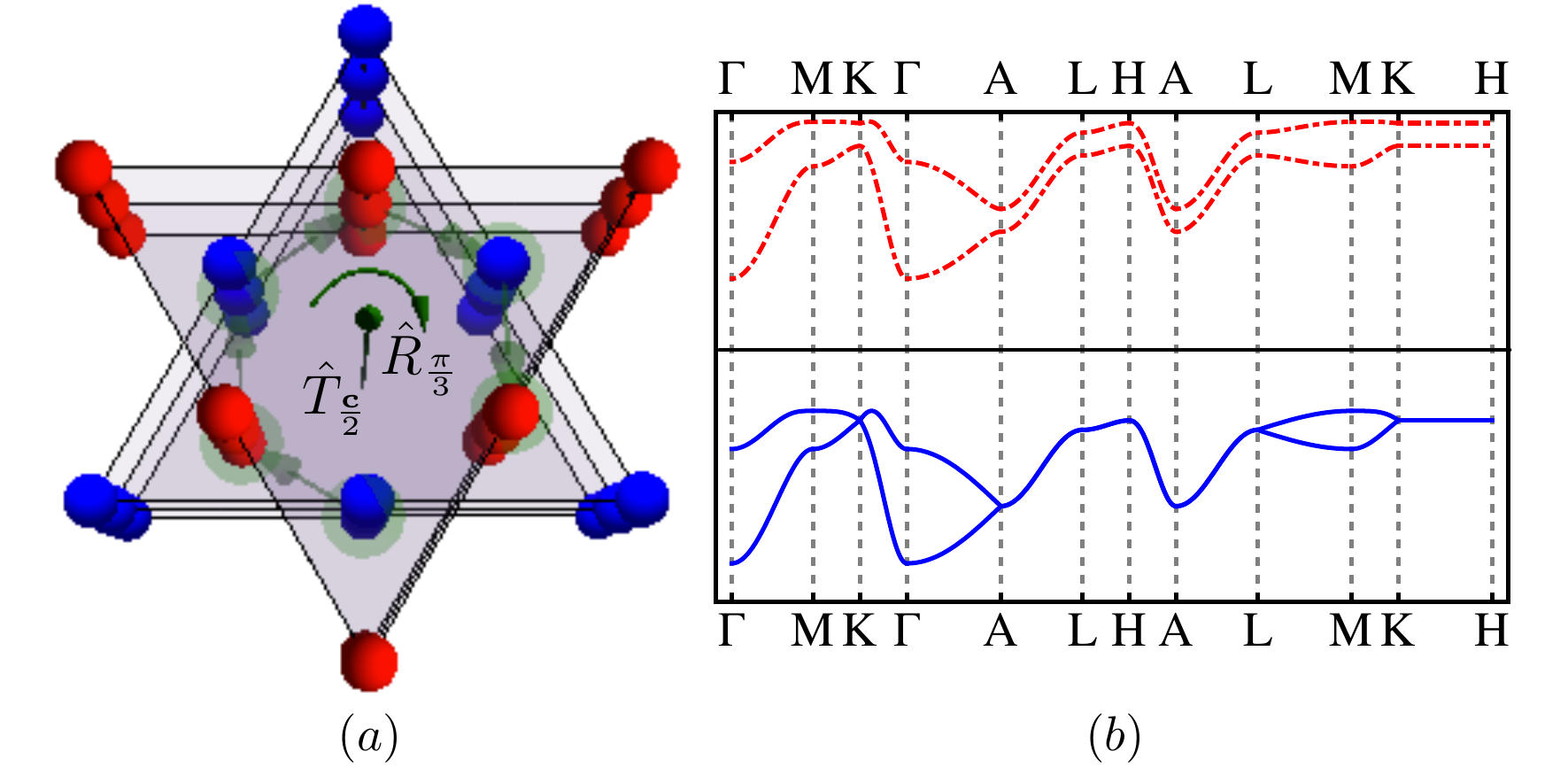}\caption{ \label{fig:hcp} Band Structure on hcp lattices. {({\bf a})} hcp crystal showing screw rotation comprised of sixfold rotation ($\hat{R}_{\frac{\pi}{3}}$) followed by half-$c$-axis translation ($\hat{T}_{\boldsymbol{c}/2}$)  out of the page; green arrows show transformation of some sites. ({\bf b}) (bottom) Tight-binding band structure (blue solid lines) showing symmetry-enforced contacts\cite{HerringTRS} between the pair of bands, leading to a minimum filling of $\nss=2$ to achieve a band insulator. (top) On breaking the screw symmetry, gaps open and allow the bands (red dot-dashed lines) to separate.}  \end{figure}
\begin{figure}
\includegraphics[width=\columnwidth]{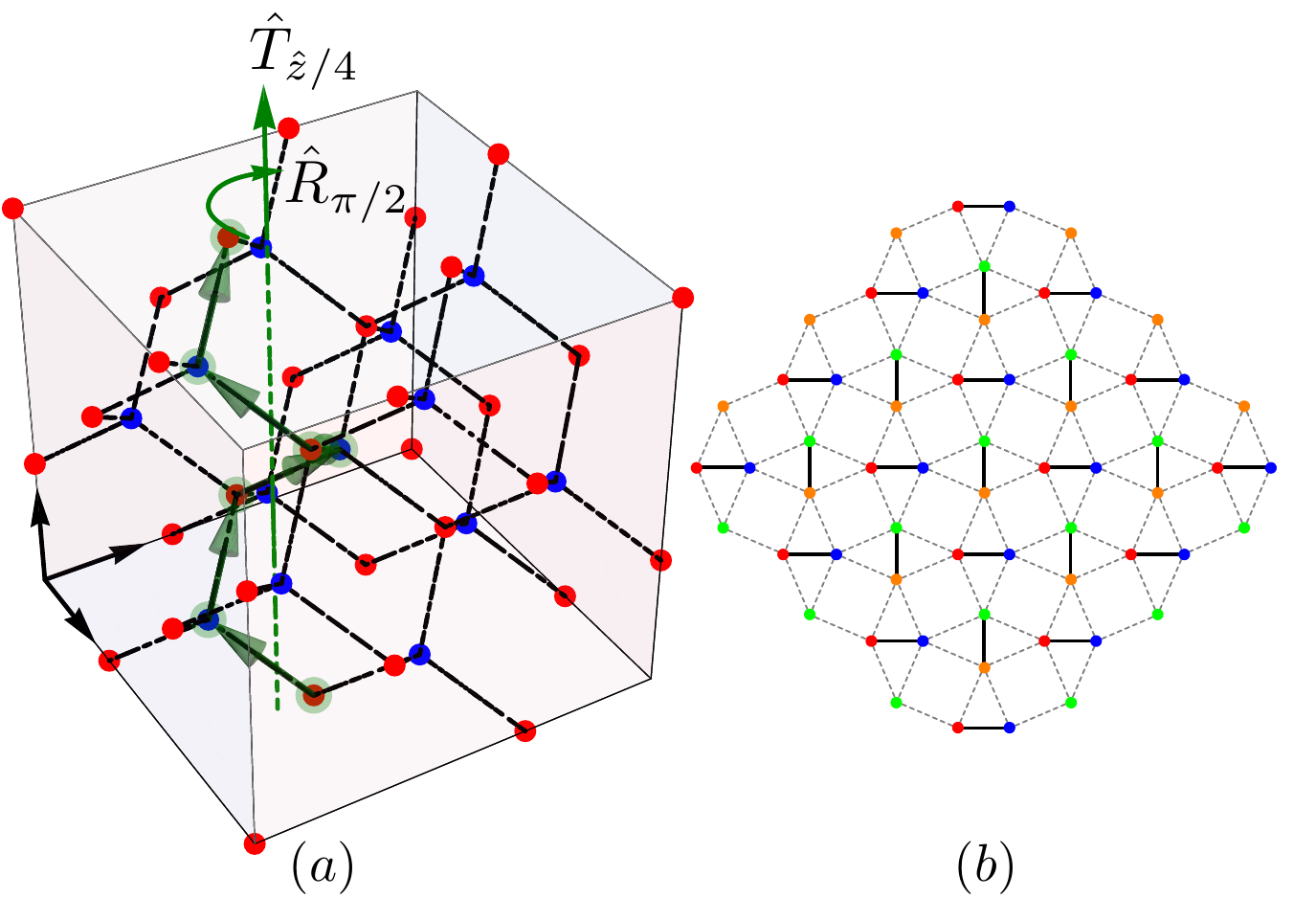}\caption{\label{fig:diamondSSL}Other Examples of Non-symmorphic Lattices. ({\bf a})  The diamond lattice  space group $Fd\bar{3}m$ has a screw rotation comprised of fourfold rotation ($\hat{R}_{{\pi}/{2}}$) followed by a quarter-translation along $\zhat$  ($\hat{T}_{\zhat/4}$).  ({\bf b}) The non-symmorphic Shastry-Sutherland lattice has the same p4g space group as the lattice of Fig.~\ref{fig:p4g}. }\end{figure}

{\it Examples.--}The statements made earlier about specific examples follow immediately from an examination of the space group symmetry.\cite{SuppMat} (i) The hcp crystal (space group $P6_3/mmc$) possesses a sixfold screw rotation about its $c$-axis, two applications of which result in a pure translation; as the only other non-symmorphic operation is a glide also of rank $2$, we conclude $\nss=2$ . A model band structure, with enforced contacts between pairs of bands is shown (Fig.~\ref{fig:hcp}); breaking the symmetry explicitly lifts these degeneracies, depicted by the broken red lines. (ii) Diamond and pyrochlore (space group  $Fd\bar{3}m$) possess a four-fold screw axis, but applying this twice yields a removable twofold screw (Fig.~\ref{fig:diamondSSL} (a)). Once again, all remaining non-symmorphic operations have rank $2$, so that $\nss=2$ for both these crystals. (iii) In a field, SCBO exhibits a magnetization plateau at half the saturation value.\cite{SebastianSSLHalfMagPlateau,SSLHalfMagExpt}  As this corresponds to $\nu=1$, unlike other plateaus which are at fractional filling,\cite{JPSJ.69.1016,Kodama11102002,AbendscheinCapponi, MilaSSL} Hastings-Oshikawa arguments cannot be applied to it. However, the two-dimensional SSL that characterizes a CuBO$_3$ layer in SCBO (Fig.~\ref{fig:diamondSSL}(b)) has the space group $p4g$ with a glide symmetry (Fig.\ref{fig:p4g}). In $d=2$ screw axes are impossible so $\nss=2$. Since $\nu=1 <\nss$, the magnetization plateau must be topologically ordered, or else break symmetry as in some proposed candidates.\cite{MomoiHalfmagtheory}

\section{Conclusions}
We have  demonstrated 
that the textbook reasoning of fillings at which band insulators may occur must be sharpened in the context of non-symmorphic latices. There, the minimum band insulating filling is not two (spinful) electrons per unit cell, 
 but a larger integer multiple.  Thus, on non-symmorphic lattices a fully symmetric insulator with two electrons per unit cell immediately implies an exotic ground state, and provides guidance to the search for novel phases of matter. 
We leave extensions of these ideas to systems with spin-orbit coupling as well as to quasicrystals, to future work.

{\bf Acknowledgements.} We thank I. Kimchi and D. Stamper-Kurn for collaboration on related work, R. Roy for many detailed conversations, M. Hermele, M. Oshikawa, S. Coh and M. Zaletel for stimulating discussions and M. Norman for valuable comments on the manuscript. This work is supported by the Simons Foundation (SAP) and by the National Science Foundation Grants No. 1066293  at the Aspen Center for Physics (SAP, DPA, AV), PHY11-25915 via the  Frustrated Magnets program at the Kavli Institute for Theoretical Physics (SAP, AV), and DMR-1007028 (DPA).

\begin{table}
\begin{center}
  \begin{tabular}{| l | c | c | c | c | }
    \hline
  d  &	 Name 			& Examples 	 & Space Group& $\nss$ \\ \hline
  2 & Shastry-Sutherland 	& SrCu$_2$(BO$_3$)$_2$ 		& $p4g$ & 2 \\
  3 & hcp    			& Be, Mg, Zn	& $P6_3/mmc$ & 2 \\
  3 & diamond 			& C, Si	 	& $Fd\bar{3}m$ &2\\ 	
  3 & pyrochlore 		& Dy$_2$Ti$_2$O$_7$ (spin ice) & $Fd\bar{3}m$ &2\\ 			
  3 & hyperkagom\'{e} 	& Na$_4$Ir$_3$O$_8$ 		& $P4_132$ & 2 \\
  3 & - 				& $\alpha$-SiO$_2$, GeO$_2$ 		& $P3_121$ & 3 \\
  3 & - 				& CrSi$_2$				& $P6_222$ & 3 \\
  3 & -	& Pr$_2$Si$_2$O$_7$,La$_2$Si$_2$O$_7$ & $P4_1$ & 4 \\
  3 & hex. perovskite  &  CsCuCl$_3$ & $P6_1$ & 6 \\
    \hline
  \end{tabular}
\end{center}\caption{\label{tab:ranks} Some non-symmorphic groups and their ranks, colloquial structure names and representative materials.}
\end{table}

\bibliographystyle{naturemag}
\bibliography{NSLSM_bib}

\begin{appendix}
\section*{Supplementary Material}
\subsection{Fourier-Space Approach, `Many-Body' Crystallography via Twist Operators and Electronic Polarization\label{app:FSC}}

In this appendix, we rederive our results (and extend them to the two exceptional cases) in a different language, that of Fourier-space crystallography. This is a reformulation of crystallography and space-group classification originally due to Bienenstock and Ewald and developed more fully by Mermin and collaborators,\citesec{Mermin:1992p1,Konig:1997p1,Konig:1999p1} that is especially well-suited to a many-body formulation. As this approach will be unfamiliar to most of our readers, we will summarize the basic ideas before discussing the many-body generalization.

A key idea of the  Fourier-space approach to crystallography is to permit all operations that preserve the {\it correlation functions} of the crystal ({\it i.e.}, its diffraction pattern) rather than its real-space positions --- as a consequence, a translation that merely shifts the origin of the crystal is equivalent to the identity. Building on this, the basic tenets of the Fourier-space approach are:\begin{enumerate}[(i)]
\item the action of a group operation on a given Fourier-space density (Bragg peak amplitude) is to multiply it by a phase factor:
\be\label{eq:symm}
 \bg: n_{\bk} \rightarrow  e^{ 2\pi i \Phi_{\bg}(\vk)} n_{g\bk}
 \ee
where $\Phi_{\bg}(\vk)$ is a linear function of $\vk$ and the $2\pi$ is introduced so that $\Phi_\bg(\vk)$ can be chosen rational.
\item a shift of real-space origin acts as a gauge transformation on the phase factors:
 \be\label{eq:gauge}
\chi:\Phi_{\bg}(\vk) \rightarrow \Phi'_\bg(\vk)= \Phi_\bg(\vk) + \chi(g\vk -\vk),\ee
 where $\chi$ is also a linear function. (Note that this is not to be confused with gauge transformations corresponding to the  conserved U(1) charge.)

\item requiring that the group multiplication rule is faithfully represented yields a group compatibility condition on the phase factors:
\be\label{eq:gcc}
\Phi_{\bg\bh}(\vk) = \Phi_\bg(\bh\vk) + \Phi_{\bh}(\vk)
\ee
\end{enumerate}
Note that in order to maintain consistency of notation with the original formulation of the Fourier-space approach, we define symmetry transformations passively,\citesec{Mermin:1992p1,Konig:1997p1,Konig:1999p1} in contrast to the main text where the transformations are active.
 The classification of different space groups then reduces to the classification of all the possible sets of phase factors, up to gauge equivalence in the sense above, given a Bravais lattice and a compatible point group. Owing to the primacy given to the diffraction pattern rather than the real-space structure, this formulation of crystallography naturally extends to quasicrystals.  As an aside, we note that in Fourier space crystallography, it is conventional to speak of `the' symmetry $g$ without worrying about the associated translation, since this emerges as the corresponding phase factor $\Phi_g(\vk)$;
 really, then, the set $\hat{G}\sim \{g|\btau\}$ in real space is replaced by $\hat{G}\sim \{g|\Phi_g\}$ in Fourier space, with the caveat that $\Phi_g$ is to be understood in a gauge-fixed sense --- this is equivalent to the statement that writing $\hat{G}\sim \{g|\btau\}$ requires a choice of origin.

A phase factor  $\Phi_{\bg}(\vk)$ for the action of  a group operation $\bg$ on a given Fourier component $n_{\vk}$ of of the density is gauge invariant if and only if the  reciprocal lattice vector $\vk$ is left invariant  by $g$ {\it i.e.} if $\bg\vk = \vk$. An essential screw or glide has $\Phi_g(\bk) \not\equiv 0$ (throughout this section, we use `$\equiv$' to mean `modulo integers', since only the fractional part of the phase function matters), and manifests itself through the vanishing of the Fourier component $n_{\vk}$ --- thus, of the 157 space groups, 155 are characterized by systematic extinctions in their Bragg patterns. The remaining two exceptional cases do not have any gauge invariant phase factors; rather, they are characterized by gauge invariant phase {\it differences}. It is evident that such phase differences correspond to an obstruction, not in fixing a gauge  to render any {\it given} screw or glide simple (free of an associated fractional translation) but in finding a gauge about which {\it all} screws/glides are simultaneously simple. Since `choosing a gauge' is Fourier-space language for `choosing a real-space origin', the result follows.  Applying symmetry to the Bloch theory of energy bands, it was shown by K\"{o}nig and Mermin that the existence of gauge invariant phase differences in space groups nos. 24 and 199 leads to a required touching of bands in the Brillouin zone, which suggests that this should be true in the interacting case as well, and which we confirm below.

The central result of this appendix --- and in some ways the most elegant formulation of our results --- is that the Fourier space approach has a natural interpretation as `many-body' crystallography if we consider the action induced by the symmetry generators on the twist operators. We shall show that in this language the phase factors appear in an analogous fashion to the single particle case,  but multiplied by the operator $\hat{\mathcal{Q}}/N$, accounting for the many-body structure; the same notion of gauge equivalence and the group compatibility condition will hold. With the usual caveats about the choice of $N$ and $\nu$, the  properties for the ground states demonstrated in the main text follow immediately. In addition, the exceptional cases appear quite straightforward in this language. This is by its nature a rather telegraphic account; we defer a detailed discussion to future work.

 We begin by introducing a full set of {\it flux-insertion operators.}
These are defined by first adiabatically turning on a gauge field $\boldsymbol{A}=\boldsymbol{k}/N$ with an integer number of flux quanta through each non-contractible loop on the torus; we call this operation $\hat{F}_{\vk}$.  Since an integer number of quanta is not observable, the final Hamiltonian is equivalent to the initial one, after applying a gauge transformation.  The necessary transformation is
 \be
 \hat{U}_{\vk} = \exp\left\{{-\frac{i}{N}\int d^d r\,  (\vk\cdot\vr) \hat{\rho}({\br})}\right\}.
 \ee
The product of the two transformations $\hat{T}_{\vk}=\hat{U}_{\vk} \hat{F}_{\vk}$, which we will colloquially term a {\it twist operator}, returns the
Hamiltonian to itself. Furthermore, since the system is assumed to be an insulator, the energy
cannot change more than an exponentially small amount.  Thus the twist operators $\{\hat{T}_{\vk}\}$ for reciprocal lattice vectors $\vk$ are operators on the space of ground states. Crucially, if the ground state is unique, it must be a simultaneous eigenstate of all of them. This will contradict the transformation properties of
the $\{\hat{T}_{\vk}\}$ in a non-symmorphic lattice if the filling is not a multiple of
the non-symmorphic rank $\mathcal{S}$.

Let us consider a symmetry operator  $\hat{G} \sim\{\bg|\btau\}$ as before. It acts on the gauge transformation operator $\hat{U}_{\vk}$ via conjugation:  a straightforward calculation yields
\be
\hat{G}^{-1} \hat{U}_{\vk} \hat{G} 					&=&\exp\left\{\frac{i}{N} \int d^d r\, (\bg\vk) \cdot\vr\hat{\rho}(\vr)\right\} e^{-i \vk\cdot\bg^{-1}\btau \hat{\mathcal{Q}}/N}
					\nonumber\\
					&\equiv& \hat{U}_{\bg\vk}\,e^{2\pi i {\Phi}_g(\vk) {\hat{\mathcal{Q}}}/{N}}
 \ee
where we have used the fact that $\vk\cdot (\bg^{-1}\vr) = g\vk\cdot\vr$, and identified the phase function
\be
{\Phi}_g(\vk) &=& \frac{1}{2\pi}\left(\bg^{-1}\btau\right)\cdot \vk =\frac{1}{2\pi} \btau\cdot (g\vk)
\label{eq:phasefunction}
\ee
which is manifestly a linear function of $\vk$. (Note that when $g\vk =\vk$ these reduce to the definitions in the main text.)

The adiabatic turning on of the flux $\hat{F}_{\vk}$ transforms
as $\hat{G}^{-1}\hat{F}_{\vk}\hat{G}=\hat{F}_{g\vk}$, and so the full flux-threading operator transforms as
\be
\hat{G}^{-1} \hat{T}_{\vk} \hat{G} = \hat{T}_{\bg\vk}\,e^{2\pi i {\Phi}_g(\vk) {\hat{\mathcal{Q}}}/{N}} \label{eq:Threadsymm}
 \ee
Furthermore, we have the relations
\begin{equation}
\hat{T}_{\vk_1}\hat{T}_{\vk_2}=\hat{T}_{\vk_1+\vk_2}.\label{eq:shifts}
\end{equation}

In order to understand the gauge transformation induced by a change of origin, consider the transformation $\vr \rightarrow\vr+\va$ which accomplishes such a change.  We then have
\be
\hat{U}_{\vk} \rightarrow \hat{U}_{\vk}\, e^{2\pi i {\chi}(\vk)  {\hat{\mathcal{Q}}}/{ N}}\label{eq:MBsymm}
\ee
where ${\chi}({\vk}) = \va\cdot\vk$ is again a linear function of $\vk$. From this, it follows that under a gauge transformation we must have $\Phi_{g}(\vk) \rightarrow \Phi'_g(\vk)$, where
\be
\Phi_{g}'(\vk) = \Phi_g(\vk) + \chi(g \vk) -\chi(\vk) \label{eq:MBgauge}
\ee
which we obtain by undoing the gauge transformation to return to the original gauge, performing the symmetry operation, and then redoing the gauge transformation to return to the new gauge.
Note that (\ref{eq:MBsymm}) has the same form as (\ref{eq:symm}), except that the phase factor is multiplied by the number operator. In addition, the phase factor transforms identically under gauge transformations --- as is evident from comparing (\ref{eq:MBgauge}) and (\ref{eq:gauge}). Going further, we can verify that the group compatibility  condition (\ref{eq:gcc}) is obeyed for a pair of transformations $\{g|\btau\}, \{h|\btau'\}$.

Thus, the `many-body' definition of the phase function that can be inferred from the action of the symmetry generators on the set of twist operators is consistent with that obtained from considering the `classical' action of symmetry on the reciprocal space amplitude of the particle density. Since the phase function, the space on which it acts ({\it i.e.}, the reciprocal lattice), the group compatibility condition and the gauge transformation all agree, the results of the Fourier-space classification of space groups  corresponds with a classification of the algebra of twist operators, with the caveat that the phase factors are multiplied by an additional factor reflecting the many-body physics.
By replacing $\hat{\mathcal{Q}}$ by its eigenvalue $N^3\nu$ (valid since particle number is a good quantum number),  it is evident that many-body phase factors differ from the phase factors for single-particle Fourier-space crystallography by a factor $N^2\nu$. For $\nu=1$ and for an appropriate choice of $N^2$ that is indivisible by any of the denominators of the nonzero ({\it i.e.}, up to addition by integers) phase factors, the fact that a phase factor or phase difference is nonzero in the single-particle case, which flags a non-symmorphic space group, retains the property that it is nonzero in the many-body sense. In this sense, the non-symmorphicity of the space group, and attendant consequences for the phase factors and the symmetry structure, carry through to the many-body case. We will demonstrate below a few consequences of this correspondence.

The proof of degeneracy of the 155 non-exceptional non-symmorphic groups is quite straightforward in the Fourier-space language. Consider the expectation value $\bra{\Psi} \hat{T}_{\vk} \ket{\Psi}$. 
 Here, $\vk$ is chosen along along the invariant plane/axis of an essential glide/screw $g$, which satisfies $\bg^q =\mathbf{1}$ (as always, $q=1,2,3,4$ or $6$). Then, $\Phi_\bg(\vk) \not\equiv 0$ by definition, and since we have $\bg\vk =\vk$ the phase is gauge invariant.  From this  combined with the fact that $\bg^q = \mathbf{1}$ and the group compatibility condition, we must have  $q \Phi_g(\vk) \equiv 0$; thus, the denominator of the phase factor must be a divisor of $q$. A further constraint is placed on the phase factor if some power $\tilde{q}<q$ of the phase factor is removable, for in this case in some gauge we must have $\tilde{q}\Phi_g(\vk) \equiv 0$, for otherwise $\bg^{\tilde{q}}$ would be essential. If we choose $\tilde{q}$ to be the smallest  power for which $\bg^{\tilde{q}}$ is removable we must have $\tilde{q}$ be a divisor of $q$ (this  can be verified by explicit computation). Thus, the denominator of $\Phi_g(\vk)$ must be $\tilde{q}$ and the numerator must be relatively prime to  this for if not, a smaller factor than $\tilde{q}$ would render $\Phi_g(\vk)\equiv 0$. From our definition of the non-symmorphic rank of a space group operation as the smallest power to which it can be raised to make it removable, it is evident that $\nss_\bg = \tilde{q}$, and from the above discussion we see that the phase factor is of the form $\Phi_g(\vk) = {p}/{\nss_\bg}$ with $p,\nss_\bg$ relatively prime. 
From this, we see that if the ground state preserves the symmetry we have  (replacing $\hat{\mathcal{Q}}$ by $N^3\nu$)
\be
\bra{\Psi} \hat{T}_{\vk} \ket{\Psi} &=&\bra{\Psi} \hat{G}^{-1}\hat{T}_{\vk} \hat{G}\ket{\Psi} \nonumber\\&=&  \bra{\Psi} \hat{T}_{g\vk} \ket{\Psi}\,e^{i 2\pi \Phi_g(\vk)N^2\nu}
\label{eq:polarizationsymmetry}
\ee
Since $g\vk = \vk$ and the gauge-invariant phase factor is given by $\Phi_g(\vk) = {p}/{\nss_\bg}$, we find that
\be\label{eq:obstacle}
\bra{\Psi} \hat{T}_{\vk} \ket{\Psi} &=&\bra{\Psi} \hat{T}_{\vk} \ket{\Psi}\,e^{i 2\pi N^2\nu p/\nss_g}
\ee

Now if the ground state $|\Psi\rangle$ is unique, then it is an eigenvector of
$\hat{T}_{\vk}$,
since this operator transforms ground states into other ground states.  It then follows that the filling must be a multiple of $\nss_g$ (by taking $N$ to be a large
number that is relatively prime to $\nss_g$). In other words, if the symmetries
are preserved, then any sufficiently large system whose size $N$
is relatively prime to $\nss_g$ must
have degenerate ground states.  To make the connection to the proof quoted in the main text apparent, observe that since $\hat{T}_{\vk} = \hat{U}_{\vk} \hat{F}_{\vk}$, and $\hat{F}_{\vk}$, $\hat{U}_{\vk}$ correspond to inserting a flux and performing a gauge transformation, it is clear that $\bra{\Psi} \hat{T}_{\vk} \ket{\Psi} = \bra{\Psi}\hat{U}_{\vk} \ket{\Psi'} =  \langle\Psi| \tilde{\Psi}\rangle$, and thus the two approaches are equivalent (modulo changes due to the active versus passive definition used in this appendix.)

This approach using twist operators can also be used to define the non-symmorphic rank arising from the interplay of all the symmetries in the group, a particular application of which is to rule out a featureless phase
for the two exceptional space groups not covered by our preceding discussion.
To do so, we first need to introduce the idea of many-body polarization. A sharp characterization of the insulating  phase is given by the electronic  polarization,\citesec{KohnInsulator,KingSmithVanderbiltPolarization} a quantity that is nonzero in the insulator but ill-defined in the metallic phase.  Specifically, in this paper we consider a definition of the polarization\citesec{Resta:2002p1, RestaPBC, RestaPolRMP}
 appropriate to the specific periodic system size $N$.
Recall that if the ground state is unique then it must be an eigenstate of all the $\hat{T}_{\vk}$. Eq. (\ref{eq:shifts}) implies
that the phase of the eigenvalues have a linear dependence on $\vk$
\be\label{eq:polarization}
 \hat{T}_{\vk}|\Psi\rangle =  e^{-{2\pi} i \vk\cdot\plz^{(N)}}\ket{\Psi}
\ee
where we use the superscript to remind us that this definition is for a finite system. (We write (\ref{eq:polarization}) in units where $e=1$; in general units, $\plz$ should be replaced by $\plz/e$ throughout.) The behavior of the finite-system polarization $\plz^{(N)}$ in the thermodynamic limit can be used to define the many-body polarization of the ground state when a Chern
number is not present.  Even in the latter case, although polarization is no longer well-defined, $\plz^{(N)}$ on a system of a fixed size still makes sense; note that as it is a vector, the group operations transform $\plz^{(N)}$. Now, replacing the expectation values in Eq. (\ref{eq:polarizationsymmetry}) by
the exponential of the polarization, and
using Eq. (\ref{eq:phasefunction}), we find that  the combination
\be
w^{(N)} &=&  -\nu N^2\btau\cdot g\vk+{\plz^{(N)}}\cdot(g\vk-\vk) \\&=&\left(-\nu N^2g^{-1}\btau+g^{-1}\plz^{(N)}-\plz^{(N)}\right)\cdot\vk\nonumber
\ee
 is an integer. Since this is true for
every reciprocal lattice vector $\vk$,
\be\label{eq:gplznrel}
g {\plz^{(N)}}+\nu N^2\btau\equiv{\plz^{(N)}}
\ee
where the equivalence is modulo a Bravais lattice vector.
Now we take (\ref{eq:gplznrel}) for systems of size $N$ and $N+1$, multiply them by $2N+3$
and $2N-1$ respectively and subtract.  This cancels off the factor of $N^2$, leaving
\be
g{\plz}+\nu \btau\equiv{\plz}
\ee
where $\plz = (2N+3) \plz^{(N)} - (2N-1) \plz^{(N+1)}$. From this, it is clear that
\be
g \left(\frac{\plz}{\nu}\right) +\btau = \frac{\plz}{\nu} + \frac{1}{\nu}\,\va_G
\ee
for some Bravais lattice vector $\va_G$.
Thus, if we define the coordinate $\br_0=\plz/\nu$ we find that
\begin{equation}
\hat{G}: \br_0\rightarrow  g\br_0 +\btau = \boldsymbol{r}_0+\frac{\boldsymbol{a}_G}{\nu}
\end{equation}
for some Bravais lattice vector $\boldsymbol{a}_G$, {\it i.e.}, $\boldsymbol{r}_0$ is invariant under the symmetries up to ${1}/{\nu^\mathrm{th}}$ of a Bravais lattice
vector. If $\nu=1$, then $\boldsymbol{r}_0$ can be used as a center for all
group operations, and hence the lattice must be symmorphic.
For a non-symmorphic lattice,
the minimum value of $\nu$ for which such a point $\boldsymbol{r}_0$ exists
can be defined as the non-symmorphic rank $\mathcal{S}$;
this is the minimum
filling where trivial insulators have a chance to exist.

Intuitively, $\plz$ is the polarization of a unit cell (relative to the origin
appearing in the definition of $\hat{U}_{\vk}$), so $\br_0={\plz}/{\nu}$
is the center of charge of the $\nu$ electrons in the cell.  If there is a single
electron per unit cell, this center of charge is well-defined (within each
unit cell).  If the group is non-symmorphic, applying the right symmetry
will change the charge-center, and thus the new
wave function is different from the original one.
When
there are more electrons in a unit cell, the center of charge becomes less well
defined, with $\nu^3$ locations in a grid in each unit cell (on account of dividing
the polarization, defined only modulo a lattice vector, by $\nu$).
Thus, for a large enough value of $\nu$, it becomes easy to ensure that these
charge centers are mapped among themselves.

This argument applies to the two exceptional groups. For any
one element of these groups,
there is a symmetric location for the charge center.  But the group elements
do not have a common point of symmetry.  We find in fact that $\nss=2$ for these groups.

 As the remaining 155 non-symmorphic space groups were covered by our previous arguments, this concludes the proof that all non-symmorphic space groups have $\nss>1$. We observe that the phase-space approach motivates two possible lines of further inquiry: (i) it extends naturally to higher dimensions, and to quasicrystals;\citesec{FSCQuasicrystals} (ii) it has a natural formulation in the language of cohomology,\citesec{FSCGroupCohomology} and suggests that the existence of gapped featureless phases with space-group symmetries may be understood within this language. Similar technology has been brought to bear, with fruitful consequences, for situations with local symmetries.\citesec{ChenGuLiuWen}

\subsection{Examples in the Main Text}

\subsubsection{Shastry-Sutherland Lattice}
The SSL is generated by the primitive vectors  $\va_1 = (b+c)\xhat, \va_2 =(b+c)\yhat$ and the four-site basis $\left\{\frac{b}{2}(\xhat-\yhat),\frac{b}{2}(-\xhat+\yhat), \frac{c}{2}(\xhat+\yhat), -\frac{c}{2}(\xhat-\yhat) \right\}$ with $b\neq c$, and corresponds to  2D space group $p4g$.  The non-symmorphic operation is a glide comprised of a reflection about a plane parallel to $\va_1$ and passing through the origin in the specified coordinate system, followed by a translation by $\va_1/2$. Since the two primitive vectors are orthogonal, it follows that this cannot be the projection of a lattice vector on the glide plane and therefore the glide is essential, leading to $\nss=2$ (in two dimensions, this is the only nontrivial value of $\nss$). Since there are four sites per unit cell, a spin-$\frac{1}{2}$ model will have saturation magnetization of $2$ per unit cell; half this value corresponds to a filling of $\nu=1$, and thus the half-magnetization plateau must either break symmetry or exhibit topological order.

\subsubsection{Hexagonal Close-Packing}
The hcp lattice is generated by the primitive vectors  $\va_1 = {a}\xhat, \va_2 = a\left(-\frac{1}{2}\xhat +\frac{\sqrt{3}}{2}\yhat\right), \va_3 = c\zhat$ and two-site basis $\left\{{\boldsymbol 0},\frac{2}{3} \va_1+\frac{1}{3}\va_2 + \frac{1}{2}\va_3  \right\}$ and corresponds to  space group $P6_3/mmc$ (no. 194). The crystal possesses a screw axis involving a sixfold rotation about $\va_3$  centered at the point $(2\va_1+\va_2)/3$, followed by a translation by $\frac{1}{2}\va_3$. Applying this twice yields the product of a pure translation (by $\va_3$) and a three-fold rotation, and so the non-symmorphic rank of this operation is $2$. The only other non-symmorphic element is a glide reflection, also of rank $2$, from which we conclude that the rank of the group is $\nss=2$.  Our result then shows that in the absence of broken lattice symmetries, a noninteracting band insulator (of spinful electrons) must be metallic unless the filling is a multiple of $4$. Furthermore, if interactions are included and open a gap, except at such fillings the resulting insulating ground state cannot be featureless, and so must break symmetry or exhibit topological order.

\subsubsection{Diamond}
 The diamond lattice is characterized by the non-symmorphic space group $Fd\bar{3}m$. In coordinates in which the fcc Bravais lattice primitive vectors are $\va_1 = \frac{a}{2}(\yhat+\zhat), \va_2 = \frac{a}{2}(\zhat+\xhat), \va_3 = \frac{a}{2}(\yhat+\zhat)$ and the two-site basis is $\left\{{\boldsymbol 0}, \frac{a}{4}(\xhat+\yhat+\zhat)\right\}$, the crystal has a $4_1$ screw axis consisting of a fourfold rotation about the $\zhat$ axis centered at $\xhat/4$, followed by a translation through $\zhat/4$. 
 Applying this twice yields a two-fold screw that is removable, as can be readily diagnosed by the fact that lattice vectors $\va_1,\va_2$ have projections of $\zhat/2$ onto the axis. As all the  remaining non-symmorphic operations are glides which also have rank $2$,  $\nss=2$. Now, take a model of local spin-$\frac{1}{2}$ moments on the diamond lattice, described by a Hamiltonian with U(1) spin symmetry and hence conserved magnetization along some axis, which we label $\hat{S}^z$. Following our prescription, we identify the charge at position $\br$ as $\hat{\mathcal{Q}}_{\br} =\hat{S}^z_{\br}+\frac{1}{2}$. From this, we find that the total charge is $\hat{\mathcal{\mathcal{Q}}}= \sum_{\br} \hat{\mathcal{Q}}_{\br} = \hat{S}^z_{\text{tot}} + N^3$, where $\hat{S}^z_{\text{tot}} = \sum_{\br} \hat{S}^z_{\br}$ and we have used the fact that there are two spins in each unit cell.  This corresponds to a filling of $\nu = s^z + 1$ where $s^z$ is the magnetization per unit cell. A gapped, featureless, and topologically trivial paramagnetic ground state (the spin analog of a featureless insulator) is ruled out unless $\nu$ is a multiple of $2$. In particular, if in addition to the U(1) symmetry the $\mathbb{Z}_2$ symmetry $\hat{S}^z_{\br}\rightarrow -\hat{S}^z_{\br}$ is preserved (which includes the fully SU(2)-symmetric limit) the total magnetization must vanish and so $\nu =1$. In this case, any gapped ground state must either break lattice symmetry, or form a topologically ordered spin liquid. This suggests a route to a three-dimensional quantum spin liquid on the diamond lattice by frustrating the tendency of spins to magnetically order. We also observe that the pyrochlore lattice has the same space group as diamond, but {\it four} sites in each unit cell; for a spin-$\frac{1}{2}$ model on this lattice with vanishing total magnetization, we find $\nu=2$ and therefore a gapped, featureless quantum paramagnet is an allowed ground state.
\end{appendix}

\bibliographystylesec{naturemag}

\bibliographysec{NSLSM_bib}

\end{document}